\documentclass[
 reprint,
superscriptaddress,
nofootinbib,
 amsmath,amssymb,
 aps,pra,
]{revtex4-1}
\usepackage{amsmath}
\usepackage{graphicx}
\usepackage{dcolumn}
\usepackage[normalem]{ulem}
\usepackage[colorlinks=true,linkcolor=blue,urlcolor=blue,citecolor=blue]{hyperref}

\begin{document}

\title{Dark resonance formation with magnetically-induced transitions: extension of spectral range and giant circular dichroism}

\author{Armen Sargsyan}
\author{Ara Tonoyan}
\author{Aram Papoyan}
\email{papoyan@ipr.sci.am}
\author{David Sarkisyan}
\affiliation{Institute for Physical Research, NAS of Armenia, Ashtarak-2, 0203 Armenia}

\date{\today}

\begin{abstract}
Dark resonances were formed via electromagnetically induced transparency for the first time involving magnetically-induced $\Delta F = \pm2$ atomic transitions of alkali metal atom, which are forbidden at zero magnetic field. The probability of these transitions undergoes rapid growth when $300 - 3000$~G magnetic field is applied, allowing formation of dark resonances, widely tunable in the GHz range. It is established that for $\Delta F = +2$ ($\Delta F =-2$) transition, the coupling laser tuned to $\Delta F = +1$ ($\Delta F =-1$) transition of the hyperfine $\Lambda$-system must be $\sigma^+$ ($\sigma^-$) polarized, manifesting anomalous circular dichroism.
\end{abstract}

\maketitle

Continuous interest to formation of dark resonances (DR) exploiting coherent population trapping (CPT) and electromagnetically induced transparency (EIT) phenomena is stipulated by a number of important applications in a variety of fields such as information storage, quantum communication, optical magnetometry, metrology, etc. \cite{1,2,3,4,5,6,7}. The dark resonances are formed on atomic transitions; meanwhile extension of their spectral range beyond zero-magnetic-field resonances could attract new applications, in particular, for tunable locking of laser radiation frequency. A straightforward way to realize this is the formation of the DR in an atomic system exposed to external magnetic field resulting in Zeeman shift of atomic transitions.
In recent years, there was a noticeable interest to $F_e-F_g = \Delta F = \pm2$ atomic transitions between the lower ($F_g$) and upper ($F_e$) levels of the hyperfine structure of alkali metals (Cs, Rb, K, Na), where $F$ is the total atomic momentum. According to the selection rules, these transitions are forbidden in a zero magnetic field, however, in a magnetic field of $\sim 1000$~G, their probabilities undergo giant increase (so called magnetically induced (MI) transitions) \cite{8,9,10,11,12}. It was demonstrated that the probabilities of MI transitions (their overall number for atoms of alkali metals is more than 70) can significantly exceed the probabilities of "allowed" atomic transitions in the range of $B = 300 - 1000$~G. Also, the following rule has been found: the intensities of MI transitions with $\Delta F = +2 (-2)$ are maximal in the case of $\sigma^+$ ($\sigma^-$) polarized excitation \cite{8,9,10,11,12}. The difference in the probabilities of some MI transitions depending on sign of the circular polarization can reach several orders of magnitude.

\begin{figure*}[tb]
\centering
\includegraphics[width=430pt]{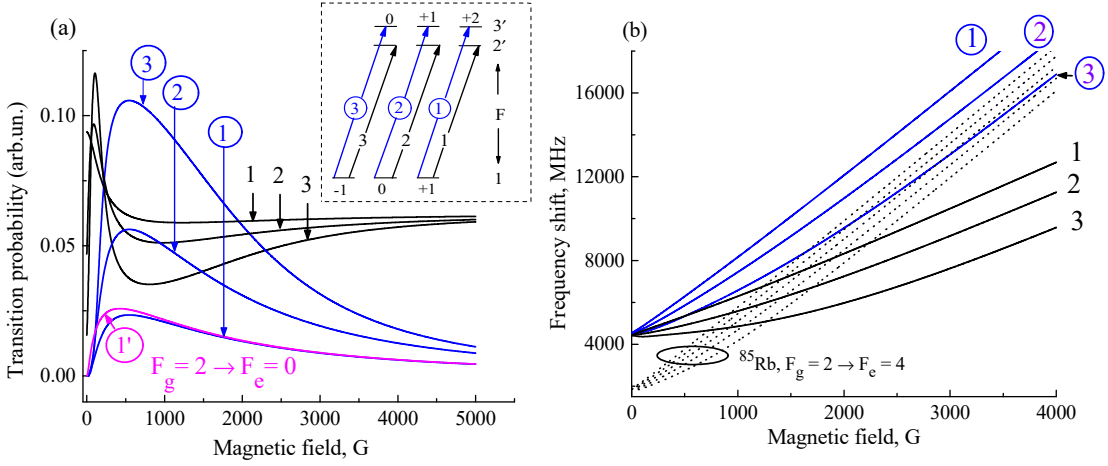}
\caption{(a) Magnetic field dependence of transition probabilities for $D_2$ line of $^{87}$Rb for $\sigma^+$ radiation. Transitions labeling is shown in the upper inset; MI transitions are marked by circled labels. {\Large \textcircled{\small 1'}} corresponds to MI transition induced by $\sigma^-$ radiation. (b) Magnetic field dependence of transition frequency shifts for $D_2$ lines of $^{87}$Rb for $\sigma^+$ radiation. Black dotted lines are for the group of five MI transitions of $^{85}$Rb $F_g=2\rightarrow F_e=4$, whose probabilities strongly decrease above $B = 1000$~G.}
\label{fig:1}
\end{figure*}

In this letter we report the first observation of dark resonances (DR) formed with MI transitions in the EIT process, realized on a $\Lambda$-system of $D_2$ line of $^{87}$Rb using MI transitions $F_g=1 \rightarrow F_e=3$ ($\Delta F =+2$) and $F_g=2 \rightarrow F_e=0$ ($\Delta F =-2$), applying $B$-field of up to 3000~G. Studies of DR were done using spectroscopic cell containing atomic vapor with a thickness of $\sim 1~\mu$m and a strong permanent magnet: thanks to the small thickness of the vapor column, the high-gradient field produced by magnet is almost uniform across the interaction region. 

For the case of  $\sigma^+$ radiation, the evolution of the probabilities of $^{87}$Rb $D_2$ line $F_g=1 \rightarrow F_e=3$ ({\Large \textcircled{\small 1}}-{\Large \textcircled{\small 3}}) MI and $F_g=1 \rightarrow F_e=2$ (labels 1-3) transitions in the magnetic field is presented in Fig.~\ref{fig:1}a.The corresponding transitions diagram is shown in the inset. The graph also contains dependence for MI transition $F_g=2 \rightarrow F_e=0$ for the case of $\sigma^-$ radiation ({\Large \textcircled{\small 1'}}). As is seen, the probability of MI transition labeled {\Large \textcircled{\small 3}} becomes dominating over all other transitions in the range of $B = 250 - 2000$~G, and remains sufficiently large up to 4000~G. Corresponding $B$-field dependences for transition frequency shifts are presented in Fig.\ref{fig:1}b. Also shifts for $^{85}$Rb $F_g=2 \rightarrow F_e= 4$ MI transitions are presented (black dotted lines), but their probabilities rapidly decrease at $B > 1000$~G \cite{12}, and that is why they do not contribute to the DR spectrum. The calculations presented in Fig.~\ref{fig:1} were done using a theoretical model that describes the modification of atomic transition probability in a magnetic field using the Hamiltonian matrix, taking into account all the transitions within the hyperfine structure. For detailed model description see e.g. \cite{8,9,10}.

\begin{figure}[tb]
\centering
\includegraphics[width=250pt]{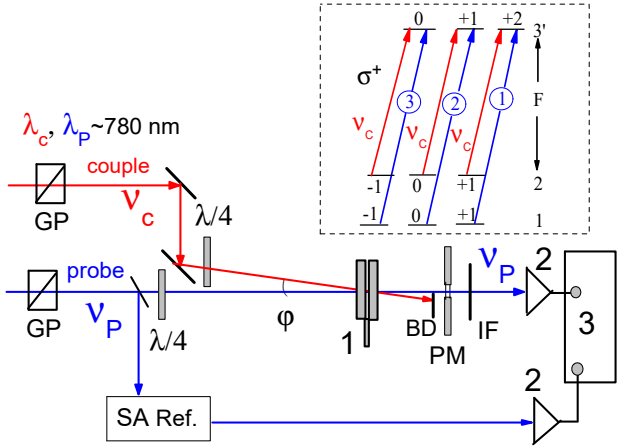}
\caption{Schematic drawing of the experimental setup. The coupling and  probe lasers ($\lambda \simeq 780$~nm) are not shown. 1 – Rb microcell in the heater; 2 – photodiodes; SA Ref. – saturated absorption (reference) unit; 3 – Tektronix TDS2014B oscilloscope; IF – 10~nm-band 780~nm interference filter; BD – beam dump;  PM – permanent ring magnet; $\varphi = 20$~mrad. Inset: the diagram of atomic levels and transitions of $^{87}$Rb $D_2$ line involved in DR formation.}
\label{fig:2}
\end{figure}

Experimental studies were carried out on a setup schematically sketched in Fig.~\ref{fig:2}. The radiation of two continuous-wave narrow-band ($\sim 1$~MHz spectral linewidth) external cavity diode lasers tunable around $\lambda = 780$~nm was used to form the coupling and the probe beams. The coupling beam frequency $\nu_C$ was fixed in resonance with $2\rightarrow 3'$ transition (hereinafter the upper levels are marked by primes), while the probe beam frequency $\nu_P$ was scanned across $1 \rightarrow 3'$ MI transition components. The energy levels and transitions for the $\Lambda$-system of $^{87}$Rb $D_2$ line relevant for the formation of DR are shown in inset of Fig.~\ref{fig:2}. The \O$\sim$1~mm coupling and probe beams were combined in a $1.5~\mu$m-thick microcell (MC) 1 containing Rb vapor (for the MC design, see \cite{13}). The temperature of the MC reservoir was kept at $110^o$C, corresponding to Rb vapor number density of $\sim 10^{13}$~cm$^{-3}$. The employed experimental configuration allowed to individually control the power of both beams ($P_C = 5-10$~mW for coupling and $P_P = 0.1-0.5$~mW for probe) and their polarizations. Glan polarizers and quarter-wave plates were used to purify the initial linear polarization of the laser beams and to form circularly-polarized beams (left-circle $\sigma^+$ or right-circle  $\sigma^-$). The coupling beam was directed onto MC at a small angle ($\varphi = 20$~mrad) with respect to the probe beam directed at normal incidence, and was blocked upon transmission by a beam dump. Part of the probe laser radiation was branched to a frequency reference (Ref) setup (saturated absorption scheme with a 3~cm-long Rb cell). The transmitted probe beam was detected by a photodiode 2 with amplifier, and recorded by a digital storage oscilloscope (3), simultaneously with the frequency reference signal.

\begin{figure}[b]
\centering
\includegraphics[width=250pt]{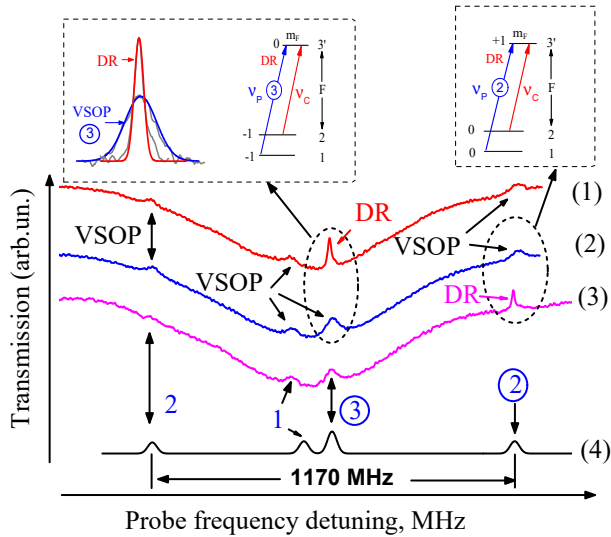}
\caption{Probe beam transmission spectra on $^{87}$Rb $D_2$ line for $L \simeq 1.5~\mu$m, $B = 720$~G. Curve (1): DR is formed when $\nu_P$ is tuned to MI transition {\Large \textcircled{\small 3}}, while the coupling frequency $\nu_C$ is set to  $2 \rightarrow 3'$ transition (see upper insets for the labeled transition diagrams). Upper-left inset: Gaussian approximations for the DR and VSOP resonances. Curve (2): the same spectrum in the absence of coupling beam. Curve (3): DR is formed when $\nu_P$ is tuned to MI transition {\Large \textcircled{\small 2}}, while the coupling frequency $\nu_C$ is set to $2 \rightarrow 3'$ transition. Curve (4): the calculated transmission spectrum of probe beam in the absence of coupling beam (compare with experimental curve (2)). The spectra are shifted vertically for clarity.}
\label{fig:3}
\end{figure}
A strong neodymium-iron-boron alloy ring-shaped permanent magnet was placed after the rear window of the MC, with the axis aligned along the probe beam propagation direction. Magnetic strength in the MC was varied by simple longitudinal displacement of the PM, calibrated using a Teslameter HT201 magnetometer. In spite of strong gradient of the magnetic field produced by a permanent magnet, the $B$-field in the atom-light interaction region can be surely considered as uniform thanks to the small thickness of MC.

The samples of recorded spectra are presented in Fig.~\ref{fig:3} for $\sigma^+$ polarized coupling and probe beams, and longitudinal magnetic field $B = 720$~G. The curve (1) shows the transmission spectrum of the probe radiation, when the coupling frequency $\nu_C$ is set to $2 \rightarrow 3'$ transition ($m_F=-1 \rightarrow m_{F'}=0$ component), for Zeeman structure of $^{87}$Rb $\Lambda$-system, see diagrams in the upper insets of Fig.~\ref{fig:3}. A dark resonance is formed when $\nu_P$ is tuned to MI transition $1 \rightarrow 3'$ ($m_F=-1 \rightarrow m_{F'}=0$ component marked as {\Large \textcircled{\small 3}}). The spectrum also contains velocity selective optical pumping (VSOP) resonances \cite{14}, which appear on atomic transition frequencies and have much larger spectral width and smaller amplitude as compared with DR. The Gaussian approximation of spectral lineshapes of DR and VSOP resonances is shown in the upper-left inset of Fig.~\ref{fig:3}. For DR, the approximation yields a 15~MHz FWHM linewidth (note that the angle between the coupling and probe beams leads to an additional spectral broadening of DR \cite{15}). Noteworthy that a pronounced DR, having a contrast of $\sim 40$~\%, is formed only when the coupling radiation is $\sigma^+$ polarized; no DR is formed for $\sigma^-$ polarized coupling. In the absence of coupling radiation (curve (2) in Fig.~\ref{fig:3}), the probe transmission spectrum contains only VSOP resonance features, both on "allowed" and MI transitions (for individual transitions labeling, see the inset of Fig.~\ref{fig:1}a). Calculated probe transmission spectrum corresponding to experimental spectrum (2) is presented by curve (4). While the coupling frequency $\nu_C$ is set to $2 \rightarrow 3'$ transition ($m_F=0 \rightarrow m_{F'}=+1$ component), the dark resonance is formed when $\nu_P$ is tuned to MI transition $1\rightarrow 3'$ ($m_F=0 \rightarrow m_{F'}=+1$ component marked as {\Large \textcircled{\small 2}}), as is seen from curve (3) of Fig.~\ref{fig:3}.

\begin{figure}[b]
\centering
\includegraphics[width=250pt]{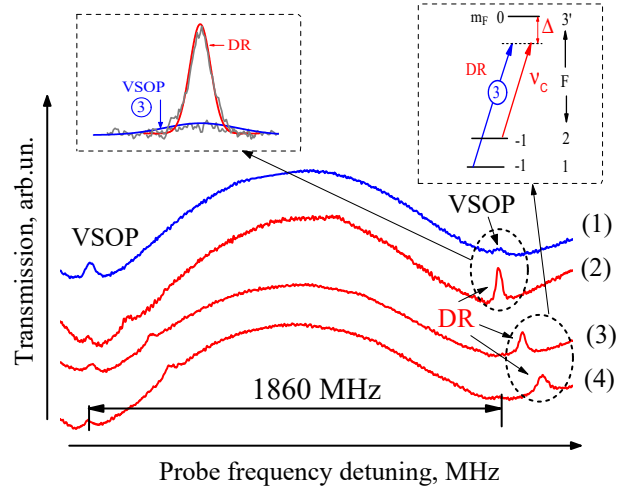}
\caption{Probe beam transmission spectra on $^{87}$Rb $D_2$ line for $L \simeq 1.5~\mu$m, $B = 2500$~G. Curve (1): no coupling beam; curve (2): DR is formed when $\nu_P$ is tuned to MI transition {\Large \textcircled{\small 3}}, while the coupling frequency $\nu_C$ is set to the $2 \rightarrow 3'$ transition. Curves (3) and (4) show the probe transmission spectra, when the detuning $\nu_C$ from the resonance with $2 \rightarrow 3'$ transition is $\Delta= 60$ and 140~MHz, respectively, exhibiting the shift of DR. The left and right insets show the DR and VSOP resonance approximations and the transitions diagram, respectively.}
\label{fig:4}
\end{figure}

As it was shown in \cite{16}, the highest contrast and the minimum linewidth of DR formed in ultrathin cells is attained when the coupling frequency $\nu_C$ is in exact resonance with the atomic transition (detuning $\Delta= 0$). In this case, the main contribution to DR formation comes from the atoms flying parallel to the cell's walls, for which the interaction time is determined by the time of flight of the atom through the laser beam: $\tau  = D/v$, where $D$ is the beam diameter, and $v$ is the thermal velocity of the atom. The situation is significantly different for the non-zero $\Delta$: in this case, the resonance occurs for atoms having velocity component $v_z =2\pi \Delta/k$ ($k=2\pi/\lambda$ is the wave number) in the direction of laser propagation $z$. The interaction time strongly decreases ($\tau = L/v_z$), thus leading to a rapid increase of the coherence dephasing rate ($L \ll D$).

The experimental results obtained with detuned frequency of coupling beam for $B = 2500$~G are presented in Fig.~\ref{fig:4}. Curves (1) and (2) show the probe transmission spectrum in the absence of the coupling beam and for the coupling radiation tuned to $2\rightarrow 3'$ transition ($\Delta= 0$), respectively (the Zeeman sublevels diagram for the related $\Lambda$-system is shown in the inset). As is seen from curve (2), a DR is recorded when the probe frequency fits the frequency of MI transition labeled {\Large \textcircled{\small 3}}. Detuning of the coupling frequency from $2 \rightarrow 3'$ transition by  $\Delta= 60$~MHz and 140~MHz (curves (3) and (4), respectively) results in contrast  reduction and broadening of DR. 

It is important to mention that when exploiting only "allowed" transitions of a $\Lambda$-system for formation of the dark resonance, the DR intensity decreases practically down to zero for $B > 1000$~G \cite{17}, while the DR formed with the MI transition labeled {\Large \textcircled{\small 3}} remains detectable even for $B = 3000$~G. We should note that DR formation in strong magnetic fields is possible by implementing a ladder-system transitions \cite{18,19}, however, in this case the key parameters (contrast and linewidth) are much worse than in the case of a $\Lambda$-system.

\begin{figure}[tb]
\centering
\includegraphics[width=250pt]{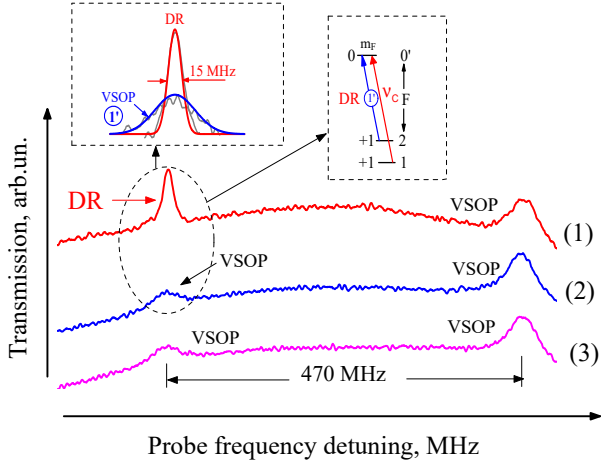}
\caption{Probe beam transmission spectra on $^{87}$Rb $D_2$ line for $L \simeq 1.5~\mu$m, $B = 570$~G. Curve (1): the coupling frequency $\nu_C$ is set to $m_F=+1 \rightarrow m_F=0$ component of $1 \rightarrow 0'$ transition (see upper-right inset for the transition diagram). DR is formed when $\nu_P$ is tuned to MI transition {\Large \textcircled{\small 1'}}. Upper-left inset: Gaussian approximations for the DR and VSOP resonances. Curve (2): the same spectrum in the absence of coupling beam. Curve (3): the coupling radiation is $\sigma^+$ polarized, no DR recorded.}
\label{fig:5}
\end{figure}

Now let us consider the results for the case when the MI transition $2 \rightarrow 0'$ ($\Delta F=-2$, marked as {\Large \textcircled{\small 1'}}) is used to form the DR (the probability of this MI versus magnetic field when using $\sigma^-$ radiation is shown in Fig.~\ref{fig:1}a). In Fig.~\ref{fig:5}, the curve (1) presents the probe transmission spectrum for $B = 570$~G. DR is recorded when both probe and coupling radiations are $\sigma^-$ polarized, $\nu_C$ is in resonance with transition $1\rightarrow 0'$, and $\nu_P$ is tuned to MI transition {\Large \textcircled{\small 1'}} (sublevels diagram for the employed $\Lambda$-system is shown in the upper-right inset). The upper-left inset shows Gaussian line approximation for a DR (fitted spectral width 15~MHz FWHM) and VSOP resonances ($\sim$45~MHz). Curve (2) shows the probe transmission spectrum in the absence of the coupling beam. No DR formed for $\sigma^+$ polarized coupling radiation (curve (3)).

Thus, it was demonstrated that in order to form a DR using MI transitions with $\Delta F = +2$, the second transition of the $\Lambda$-system should be excited by a $\sigma^+$ polarized coupling radiation, while in the case of a MI transitions with $\Delta F = -2$, the coupling radiation acting on the second transition should be $\sigma^-$ polarized. This different behavior is an evidence for magnetically-induced circular dichroism (MCD) \cite{11,12,20}. As a quantitative figure of merit for MCD, we can introduce a coefficient $C_{MCD} = (I_{\sigma^+}- I_{\sigma^-})/(I_{\sigma^+}+I_{\sigma^-}$), where $I_{\sigma^+}$ and $I_{\sigma^-}$ are the transition intensities for the coupling radiation with polarizations $\sigma^+$ and $\sigma^-$, respectively. It is easy to see that the sign of $C_{MCD}$ indicates prevailing contribution to the achieved DR intensity (the sign is positive for $\sigma^+$ and negative for $\sigma^-$ coupling).

\begin{figure}[tb]
\centering
\includegraphics[width=250pt]{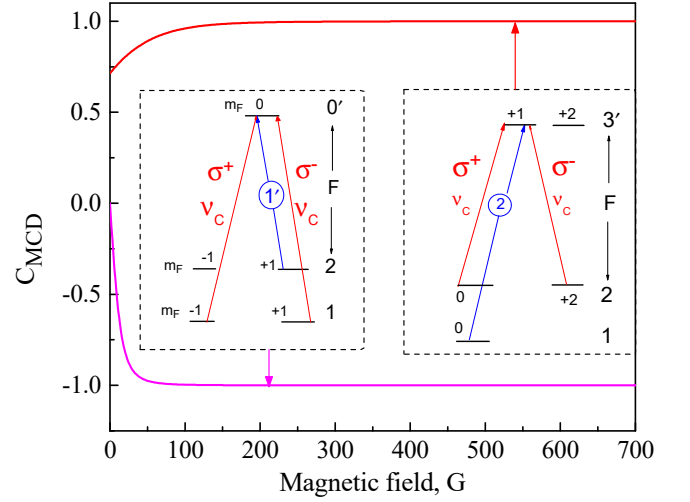}
\caption{Calculated dependence of $C_{MCD}$ coefficient on the magnetic field for the transition configurations shown in the insets (see the text).}
\label{fig:6}
\end{figure}

Figure \ref{fig:6} shows the $B$-field dependence of $C_{MCD}$ coefficient calculated using the theoretical model described in \cite{8,9,10}. The calculations were done for two cases. 1) Coupling radiation $\nu_C$ is in resonance with $1 \rightarrow 0'$ transition, and the probe frequency $\nu_P$ is tuned to MI transition {\Large \textcircled{\small 1'}}; 2) Coupling radiation $\nu_C$ is in resonance with $2\rightarrow 3'$ transition, and the probe frequency $\nu_P$ is tuned to MI transition {\Large \textcircled{\small 2}}. Both $\sigma^+$ and $\sigma^-$ polarizations are considered (the Zeeman sublevels of the $\Lambda$-system are shown in the left and right insets in Fig.~\ref{fig:6}, respectively). As one can see, the first case (the lower curve in Fig.~\ref{fig:6}), the probability of DR formation with $\sigma^+$ polarized coupling radiation becomes zero ($C_{MCD} = -1$) already for $B > 50$~G, so that only $\sigma^-$ polarized coupling beam leads to the formation of DR, which is consistent with the experiment. In the second case (the upper curve in Fig.~\ref{fig:6}), the probability of DR formation with $\sigma^-$ polarized coupling radiation becomes zero ($C_{MCD} = +1$) for $B > 100$~G, so that only $\sigma^+$ polarized coupling beam leads to formation of the DR, that is also confirmed by the experiment. We should note that in the second case of these calculations we considered MI transition marked {\Large \textcircled{\small 2}} but not {\Large \textcircled{\small 3}}, since for the latter the probability for coupling radiation is 1.4 times less. However, this is insignificant, affecting only the $B$-field value for which condition $C_{MCD} = +1$ is achieved ($B > 150$~G). Noteworthy mentioning that it is not possible to predict a priori the observed MCD behavior, as DR formation occurs for the different ground levels linked by coupling and probe radiations.

As the calculations show, the above described rule for coupling beam polarization holds valid for formation of DR also in $D_2$ line systems of other alkali atoms: $^{133}$Cs (24 MI transitions), $^{85}$Rb (16 MI), $^{39}$K (8 MI), $^{41}$K (8 MI), and $^{23}$Na (8 MI). Thus, the following rule manifesting abnormal circular dichroism is established for DR formation: when using the MI transitions with $\Delta F = +2 (-2)$, the second transition of the $\Lambda$-system $F_e-F_g = \Delta F = +1 (-1)$ should interact with $\sigma^+$ ($\sigma^-$) polarized coupling radiation.

Formation of dark resonances with MI transitions is advantageous for shifting narrow resonances far away from zero-magnetic-field atomic transitions. Particularly, the frequency of MI {\Large \textcircled{\small 3}} is shifted by 9~GHz for $B \sim 3000$~G (Fig.~\ref{fig:1}b). Formation of narrow optical resonances with strongly shifted frequencies allows extending the spectral range of stabilization of diode laser radiation frequency via locking to atomic transitions \cite{21}. Much narrower spectral width of DR can be attained by using centimeter-long cells with anti-relaxation coating, and employing coherently-coupled probe and coupling radiations derived from a single laser beam \cite{1,2}.

\section*{Acknowledgments}
The authors thank to J. Goldwin and G. Hakhumyan for useful discussions. The work was funded by the State Committee of Science, MES of Armenia (research project No. 18T-1C018).

\end{document}